\documentclass[3p,times]{elsarticle}

\usepackage{ecrc}
\usepackage{hyperref}

\volume{00}

\firstpage{1}

\journalname{Nuclear Physics A}

\runauth{Yuri Kharlov for the ALICE Collaboration}

\jid{nupha}

\usepackage{amssymb}

\usepackage[figuresright]{rotating}

\newcommand{\sqsNN}{\sqrt{s_{_{NN}}}}
\newcommand{\pT}{\ensuremath{p_{\rm T}}} 

\begin{document}

\begin{frontmatter}



\dochead{}

\title{Neutral meson production in pp and Pb-Pb collisions at LHC}


\author{Yuri Kharlov for the ALICE Collaboration}

\address{Institute for High Energy Physics, Protvino 142281, Russia}

\begin{abstract}
  The ALICE detector at the LHC studies $\pi^0$ and $\eta$ meson
  production by two complementary methods, using electromagnetic
  calorimeters and the central tracking system for converted
  photons. Production spectra of $\pi^0$ and $\eta$ mesons were
  measured with ALICE in pp collisions at LHC energies at mid-rapidity
  in a wide transverse momentum range. The spectrum and the nuclear
  modification factor $R_{AA}$ of $\pi^0$ measured in Pb-Pb collisions
  at different centralities, show a clear pattern of strong
  suppression in a hot QCD medium with respect to pp collisions.
\end{abstract}


\end{frontmatter}


\section{Introduction}
\label{sec:intro}

Nuclear matter created in heavy-ion collisions at LHC energies is
characterized by energy density which is sufficiently large to achieve
a state of deconfined quarks and gluons. Partons produced in initial
hard processes, traverse this dense matter and loose their energy via
radiative or collisional mechanisms. Measurememnt of the energy loss
by hard-scattered partons escaping the medium is a direct probe to the
energy density of the medium. The most straightforward access to the
parton energy loss is a measurement of nuclear modification of
inclusive identified hadron spectra. Light neutral pions and $\eta$
mesons are among those hadrons which production can be calculated
within perturbative Quantum Chromodynamics, as their fragmentation
functions are well known from the bulk of experimental data
\cite{Sassot:2010bh,Aidala:2010bn}. Detection of $\pi^0$ and $\eta$
mesons is performed via two-photon decays and identification of these
mesons is possible in a wide transverse momentum range.

Parton energy loss in a dense QCD medium is measured via comparison of
hadron production in heavy-ion collisions of different centralities
and the relevant hadron spectra in pp collisions.  Identified hadron
production studies in pp collisions are thus an essential part of a
heavy-ion physics programme, however not exclusively interesting as a
reference for heavy-ion collisions.  Hadron spectra measured in pp
collisions are important to tune and validate theoretical predictions
and models on hadron productions. Most of these models were proven for
a good description of particle spectra in pp and $\rm{p}\bar{\rm p}$
collisions up to the energy of the Tevatron collider,
$\sqrt{s}=1.8$~TeV. Extrapolation of the model parameters to the LHC
energy $\sqrt{s}=7$~TeV and comparison of new experimental data with
model calculations provides a crucial test of predictive power of the
models.

We present measurements of $\pi^0$ and $\eta$ mesons in pp 
collisions at LHC energies $\sqrt{s}=0.9$, $2.76$, $7$~TeV 
and $\pi^0$ in Pb-Pb collisions at $\sqsNN=2.76$~TeV with the
ALICE detector.  The measured spectra are compared with pQCD
calculations.  The measurement of the nuclear modification factor
$R_{AA}$ for $\pi^0$'s in Pb-Pb collisions in four centrality bins is
also presented.

\section{Detector and data sample}
\label{sec:detector}

\begin{figure}[ht]
  \centering
  \includegraphics[width=0.30\hsize]{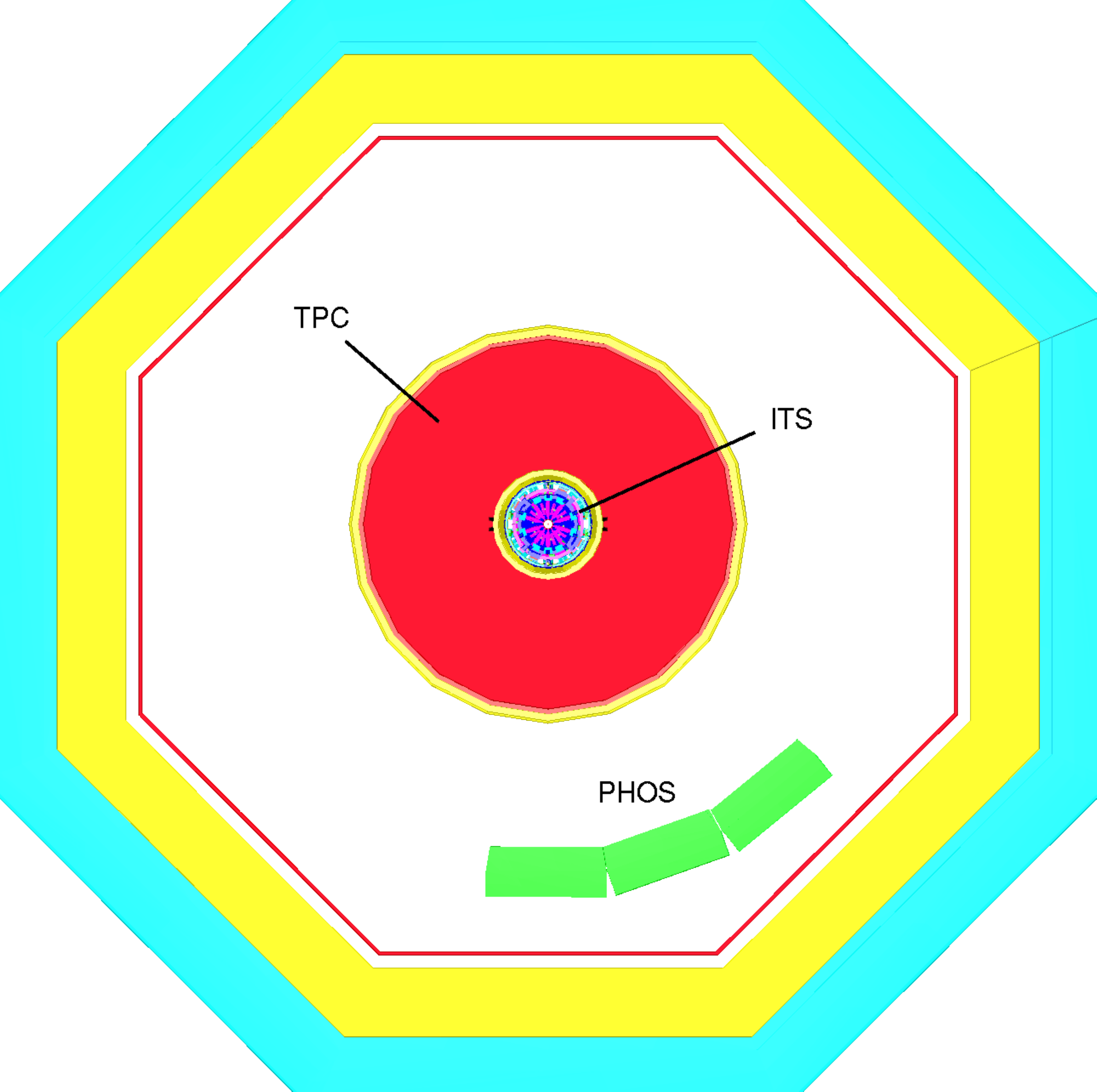}
  \caption{ALICE detectors for photon reconstruction, the Photon
    Spectrometer and the central tracking system consisting of ITS and
    TPC.}
  \label{fig:ALICE-setup}
\end{figure}

Neutral pions and $\eta$ mesons were measured with the ALICE
experiment via their two-photon decay channel. Two alternative and
complementary methods were employed in this analysis. In one method,
photons from $\pi^0$ and $\eta$ decays were measured with the Photon
Spectrometer PHOS which is a precise high-granularity electromagnetic
calorimeter built of lead tungstate crystals
\cite{Dellacasa:1999kd}. It consists of 3 modules with $64\times 56$
crystals packed into a rectangular matrix in each module. The
acceptance of the PHOS spectrometer is $60^\circ$ in azimuthal angle
and $|\eta|<0.13$ in pseudorapidity. Another method of photon
reconstruction was based on identification of photons converted to
$e^+e^-$ pairs in the medium of the inner ALICE detectors. This
analysis is referred below to as a photon conversion method (PCM).
Conversion electrons and positions were reconstructed by the ALICE
central tracking system which consists of the Inner Tracking System
(ITS) \cite{Aamodt:2010aa} and the Time Projection Chamber (TPC)
\cite{Alme:2010ke}. The central tracking system has an acceptance of
the azimuthal angle $360^\circ$ and the pseudorapidity range
$|\eta|<0.9$.  Photons can be reconstructed from $e^+e^-$ pairs if
conversion takes place between the beam interaction point and the half
radius of the TPC, $R<180$~cm with an integrated material budget as
$(0.114 \pm 0.005)X_0$. Low probability of photon conversion is
compensated by the wide acceptance of the central tracking system,
making the overall neutral meson reconstruction efficiency comparable
to that of the PHOS detector.  See Fig.\ref{fig:ALICE-setup} for a
schematic view of the ALICE detectors used for photon
reconstruction. Additional information can be found in
\cite{Abelev:2012cn}.

Data used for the neutral meson measurements in pp collisions were
recorded in 2010--2011 with the minimum bias trigger
\cite{Aamodt:2010ft} which selected most of pp inelastic events with
efficiencies $0.91^{+0.03}_{-0.01}$, $0.88^{+0.06}_{-0.04}$ and
$0.85^{+0.06}_{-0.04}$ for the three collision energies $\sqrt{s}=0.9$,
$2.76$ and $7$~TeV, respectively. Integrated luminosities of the
analyzed pp data samples are compiled in Table~\ref{tab:intlumi}.
\begin{table}[h!]
  \centering
  \begin{tabular}{|l|p{0.1\hsize}|p{0.1\hsize}|p{0.1\hsize}|} \hline
           & \multicolumn{3}{c|}{$\sqrt{s}$ (TeV)}  \\ \cline{2-4}
    Method & \hfil 0.9  \hfil & \hfil 2.76 \hfil & \hfil 7 \hfil \\ \hline\hline
    PCM    & \hfil 0.14 \hfil & \hfil 1.05 \hfil & \hfil 5.6 \hfil \\ \hline
    PHOS   & \hfil 0.14 \hfil & \hfil 0.63 \hfil & \hfil 5.7 \hfil \\ \hline
  \end{tabular}
  \caption{Integrated luminosities $\cal L_{\rm int}$ in nb$^{-1}$ of
      the analyzed pp data samples.}
  \label{tab:intlumi}
\end{table}
Pb-Pb collision results presented here were obtained from data
collected by ALICE in 2010. The minimum bias trigger in Pb-Pb
collisions was fully efficient, and the integrated luminosity used in
the analysis was $2~\mu{\rm b}^{-1}$.

\section{$\pi^0$ and $\eta$ meson spectra in pp collisions}
\label{sec:ppSpectra}

\begin{figure}[ht]
  \includegraphics[width=0.48\hsize]{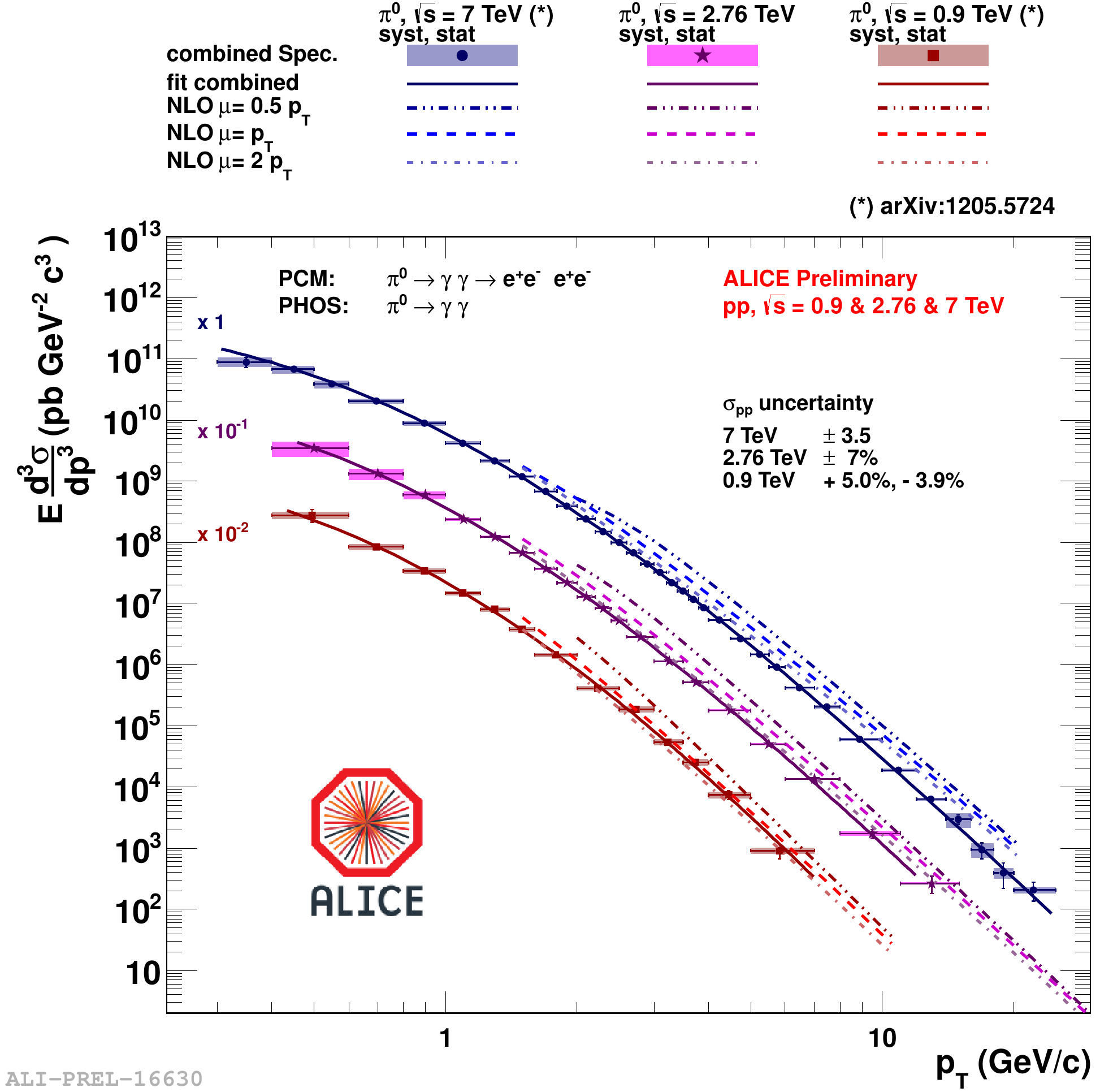}
  \hfil
  \includegraphics[width=0.48\hsize]{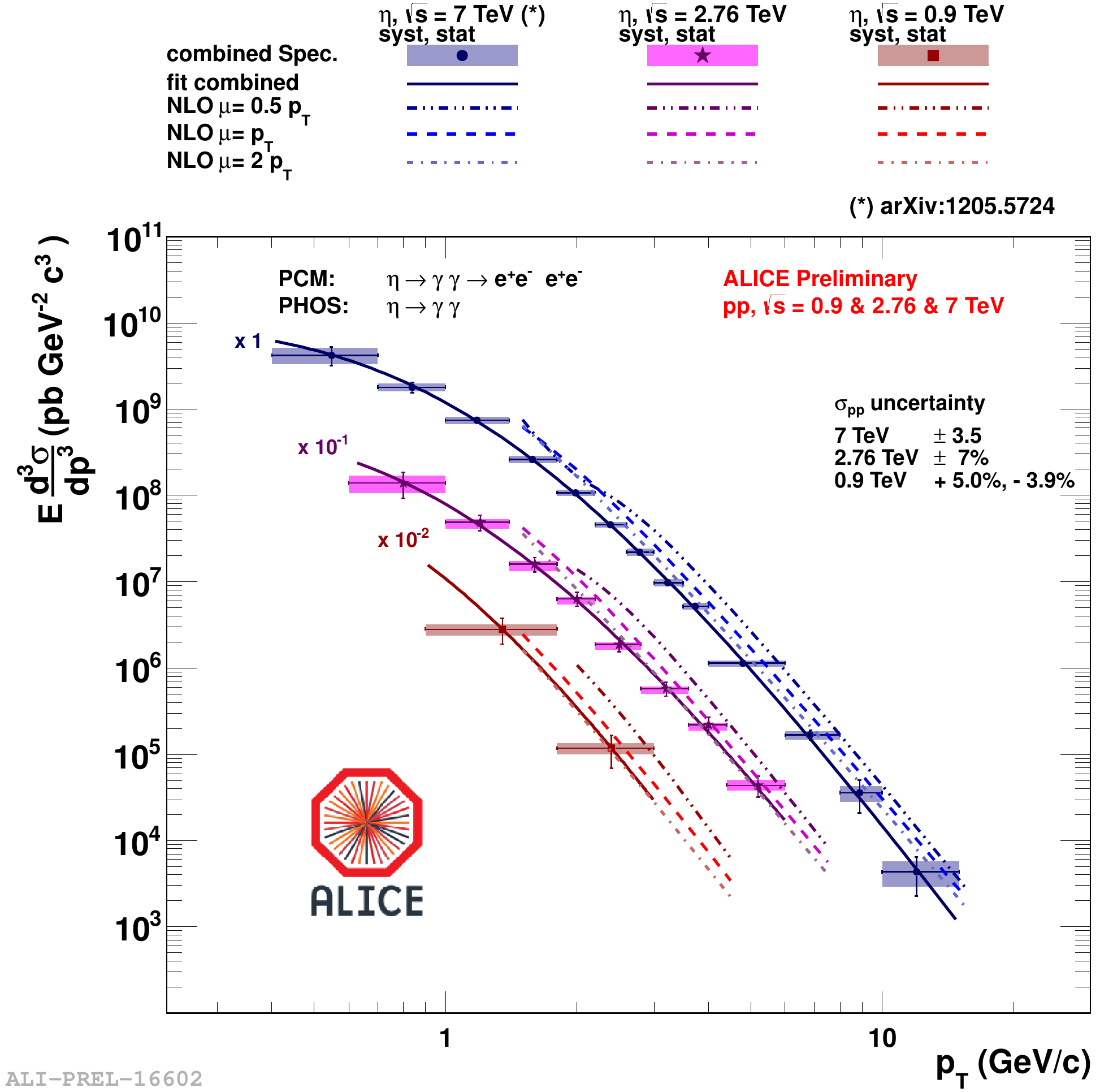}
  \caption{Differential cross sections of $\pi^0$ and $\eta$ meson
    production in pp collisions at $\sqrt{s}=0.9, 2.76$ and $7$~TeV
    compared to NLO pQCD calculations with different QCD scales, shown
    by curves. Statistical and systematic uncertainties are shown by
    solid lines and shaded area respectively.}
  \label{fig:pp-xsec}
\end{figure}

Production spectra of $\pi^0$ and $\eta$ mesons in pp collisions were
measured by ALICE at all three collision energies
\cite{Abelev:2012cn}, \cite{Reygers:2011ny} and are shown in
Fig.\ref{fig:pp-xsec}. Wide $\pT$ ranges and fairly good accuracy of
the measured differential cross sections were considered as a valuable
experimental input to test pQCD calculations at the LHC
energies. Next-to-Leading Order (NLO) pQCD calculations of $\pi^0$ and
$\eta$ meson production with a set of parton distribution functions
CTEQ6M5 and different sets of fragmentation functions, DSS, BKK and
AESSS were performed and compared with the ALICE measurements.
The pQCD calculations describe the data at $\sqrt{s}=0.9$~TeV well,
however larger center-of-mass energies are overestimated by the
calculations by a factor of $2-3$. Additionally, the calculated
\pT\ spectra show a harder slope compared to the measured results.
This discrepancy of pQCD and data can be explained by uncertainties in
fragmentation functions, especially in the gluon fragmentation
function, because pion production is considered to be dominated by
gluon fragmentation at LHC energies in a wide $\pT$ range, up to
100~GeV/$c$ \cite{Sassot:2010bh,Chiappetta:1992uh}.

\begin{figure}[ht]
  \parbox[t]{0.48\hsize}{
    \includegraphics[width=\hsize]{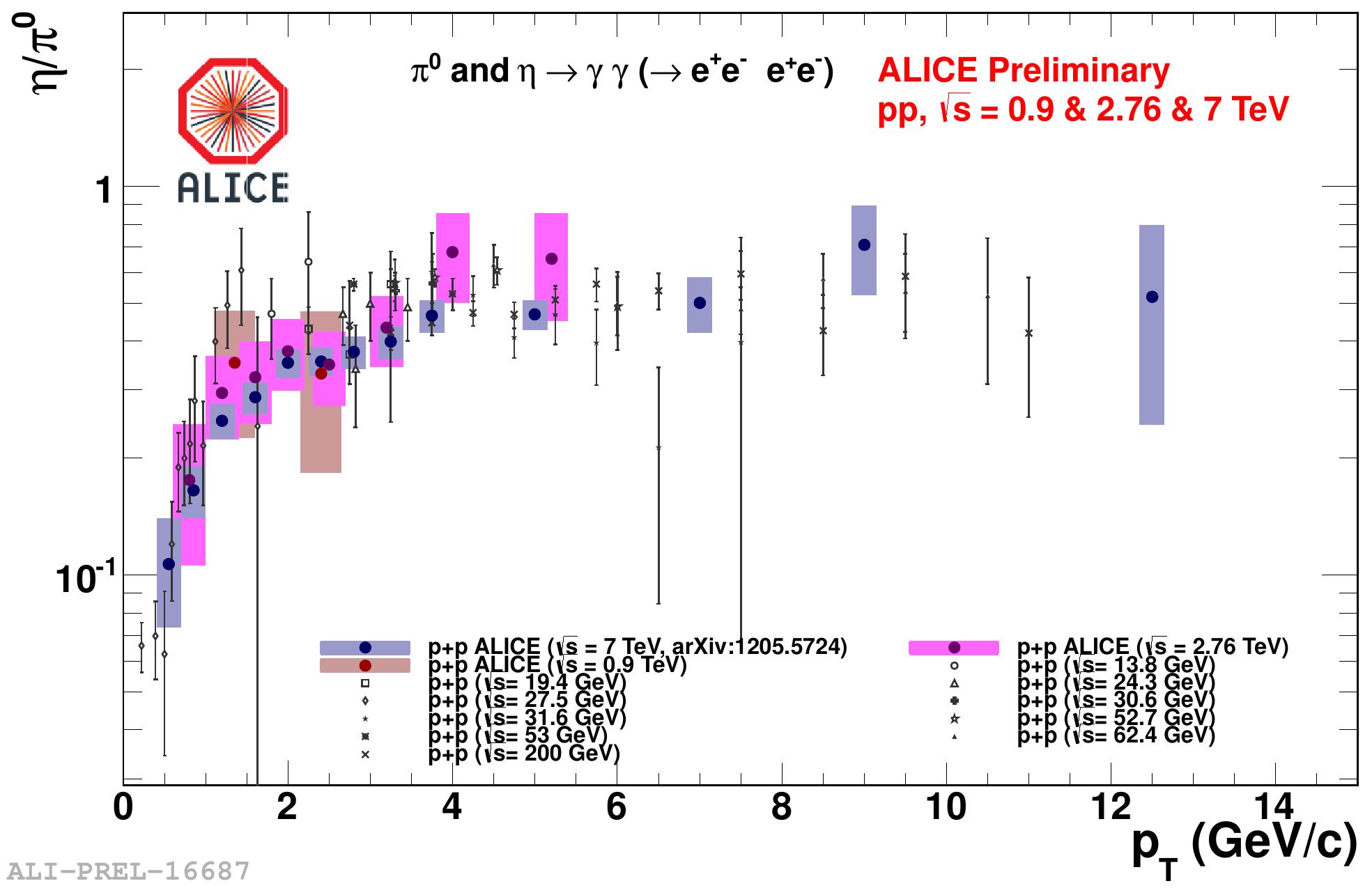}
    \caption{World compilation of the $\eta$ to $\pi^0$ production
      spectra ratio in pp collisions at low energies, along with the ALICE
      results at $\sqrt{s}=0.9$, $2.76$ and $7$~TeV, shown with the
      combined statistical and systematic uncertainties.}
    \label{fig:pp-eta2pi0}
  }
  \hfill
  \parbox[t]{0.48\hsize}{
    \includegraphics[width=\hsize]{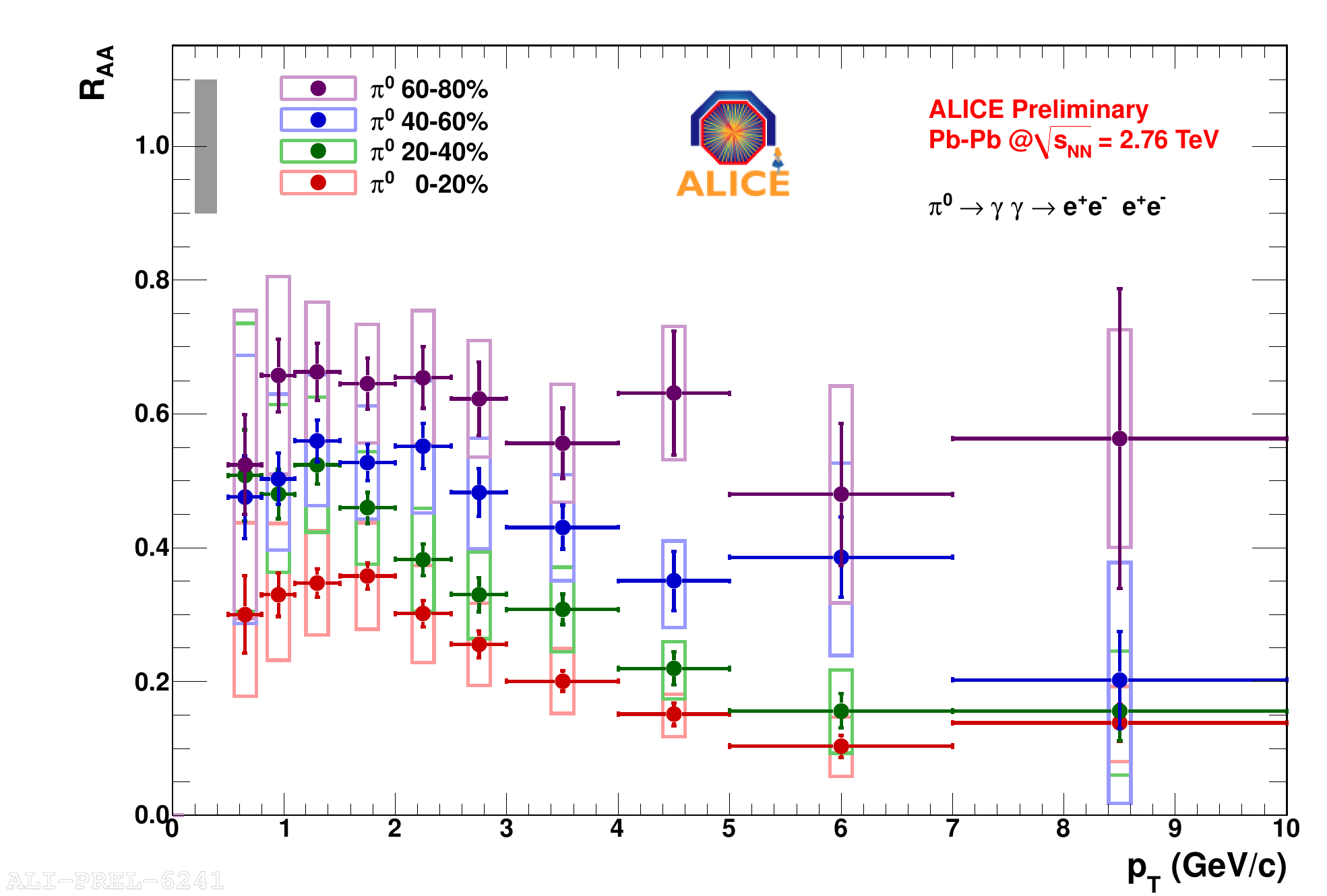}
    \caption{Nuclear modification factor $R_{AA}$ for $\pi^0$
      production in Pb-Pb collisions at $\sqsNN=2.76$~TeV in four
      centrality classes. Statistical and systematic uncertainties are
      shown by solid lines and boxes respectively.}
    \label{fig:pp-pi0RAA}
  }
\end{figure}

The ALICE results of the $\pi^0$ and $\eta$ spectra in pp at
$\sqrt{s}=0.9$, $2.76$ and $7$~TeV allowed to measure the ratio
$\eta/\pi^0$ vs \pT.  The comparison of the ALICE results with the
compilation of $\eta/\pi^0$ ratio measured by many lower-energy
experiments, presented in Fig.\ref{fig:pp-eta2pi0}, shows an universal
behavior at any collision energy.  The ratio $\eta/\pi^0$ grows up to
$\pT=2-3$~GeV/$c$ and then saturates at the same value $\pi^0/\eta
\approx 0.5$. As it was shown in \cite{Abelev:2012cn}, NLO pQCD
calculations of this ratio is consistent with data in spite of
discrepancy of individual $\pi^0$ and $\eta$ spectra.
%

\section{$\pi^0$ spectrum in Pb-Pb collisions}
\label{sec:PbPbSpectra}

The available statistics of Pb-Pb collisions taken by the ALICE
experiment in 2010 allowed to split data to 4 collision centrality
classes, $0-20\%$, $20-40\%$, $40-60\%$ and $60-80\%$
\cite{Collaboration:2011rta}. Both $\pi^0$ reconstruction methods, PCM
and PHOS, were employed in Pb-Pb collisions. High-multiplicity
environment in heavy-ion collisions affects significantly the
performance of both methods to extract the $\pi^0$ signal. In PHOS,
the overall detector occupancy in the most central collisions reaches
up to $20\%$ of the total number of cells which makes the efficiency
calculation challenging. In PCM, an increase of the overall track
multiplicity in central Pb-Pb collisions by a factor of 400
\cite{Aamodt:2010cz} in comparison with that in pp collisions leads
also to a high combinatorial background of the converted photon
candidate pairs. Here we present the $\pi^0$ spectrum measurement by
the photon conversion method only \cite{ConesaBalbastre:2011zi}.

The nuclear modification factor of the $\pi^0$ production $R_{AA}$ was
defined as a ratio of the $\pi^0$ spectrum measured in Pb-Pb collision
of a given centrality and normalized to the number of binary
nucleon-nucleon collisions $N_{\rm bc}$ \cite{Collaboration:2011rta},
to the spectrum in pp collisions derived to inelastic events. The
result shown in Fig.\ref{fig:pp-pi0RAA} demonstrates a strong
dependence of the $\pi^0$ suppression on collision centrality.  It is
instructive to note that the \pT\ dependence of $R_{AA}$ in the most
central collisions reveals the same behavior for most hadron spectra
measured at the LHC energy, see e.g. a review
\cite{Christiansen-HP2012} in these proceedings.

\section{Conclusion}
\label{sec:Conclusion}

The $\pi^0$ and $\eta$ spectra and the nuclear modification factor
$R_{AA}$ measurements have been performed with the ALICE detector. The
measured spectra in pp collisions at $\sqrt{s}=0.9$~TeV agree with
pQCD calculations, but show disagreement with that at $\sqrt{s}=2.76$
and $7$~TeV. The nuclear suppression measured by ALICE has a strong
dependence on the collision centrality and it increases with collision
energy in comparison with RHIC \cite{Adare:2008qa}, indicating an
increase of the parton energy loss with energy.

This work was partially supported by the grants RFBR 10-02-91052 and
RFBR 12-02-91527. 



\bibliographystyle{elsarticle-num}
\bibliography{pi0raa.bib}







\end{document}